\documentclass[aps,prl,twocolumn,superscriptaddress,nopacs,amsmath,amssymb,final,letterpaper,longbibliography]{revtex4-1}

\usepackage[utf8]{inputenc}
\usepackage{calc}
\usepackage{graphicx}
\usepackage{amsmath,amssymb,amsthm}
\usepackage{txfonts}
\usepackage{bm}
\usepackage{color}
\usepackage[hidelinks]{hyperref}

\newcommand{\emid}{\epsilon_{\textrm{MID}}}
\newcommand{\eend}{\epsilon_{\textrm{END}}}
\newcommand{\dalg}{d_{\textrm{ALG}}}
\newcommand{\deven}{d_\textrm{EVEN}}
\newcommand{\dmdl}{d_\textrm{MDL}}

\begin{document}
\author{Andrea J. \surname{Allen}}
\affiliation{Vermont Complex Systems Center, University of Vermont, Burlington VT, USA}
\affiliation{Applied Clinical Research Center, Children’s Hospital of Philadelphia, Philadelphia, PA, USA}
\author{Cristopher \surname{Moore}}
\affiliation{Santa Fe Institute, Santa Fe NM 87501, USA}
\author{Laurent \surname{H\'ebert-Dufresne}}
\affiliation{Vermont Complex Systems Center, University of Vermont, Burlington VT, USA}
\affiliation{Department of Computer Science, University of Vermont, Burlington VT}

\title{Compressing the chronology of a temporal network with graph commutators}

\begin{abstract}
Studies of dynamics on temporal networks often represent the network as a series of ``snapshots,'' static networks active for short durations of time. We argue that successive snapshots can be aggregated if doing so has little effect on the overlying dynamics. We propose a method to compress network chronologies by progressively combining pairs of snapshots whose matrix commutators have the smallest dynamical effect. We apply this method to epidemic modeling on real contact tracing data and find that it allows for significant compression while remaining faithful to the epidemic dynamics.
\end{abstract}

\maketitle

\paragraph{\textbf{Introduction}}
High resolution temporal interaction data is now simple to obtain and widely available thanks to methods such as radio frequency identification~\cite{cattuto2010dynamics} or Bluetooth signals~\cite{sapiezynski2019interaction}. Temporal interactions have rich dynamics in continuous time, yet we often want to combine intervals of temporal data into a series of simpler, static networks in order to compress the data, reduce analytical complexity, or streamline data collection efforts. For example, digital contact tracing protocols ping devices at fixed intervals to save energy and lighten data requirements.
However, it is nontrivial to determine when and how to aggregate temporal data without losing critical information about the dynamics of the interactions. 

Many methods currently exist to represent and analyze temporal networks~\cite{holme_temporal_2012}, and to find patterns in network structure and dynamics. This includes algorithms for detecting temporal states~\cite{masuda_detecting_2019}, dynamical approaches for generating synthetic temporal network data~\cite{peixoto_modelling_2017,Zhang2017}, tools to identify community structure in time-varying networks~\cite{ghasemian_detectability_2016}, data-driven approaches to model dynamics on temporal networks by determining change points~\cite{p_peixoto_change_2018}, and methods to represent key temporal features as static networks \cite{holme_epidemiologically_2013, scholtes_higher-order_2016}. The dynamics of epidemic spread on temporal networks is well-studied \cite{rocha_simulated_2011, genois_compensating_2015, ren_epidemic_2014, valdano_analytical_2015}, as are synchronization \cite{boccaletti2006synchronization,stilwell2006sufficient,zhang2021designing} and control dynamics \cite{li2017fundamental}.

A continuing challenge is the interplay of dynamics \textit{on} the network (changes in variables on nodes and edges) with dynamics \textit{of} the network (where the topological structure changes over time). When the timescales of these two are well separated, we can take one of two limits. If the dynamics on the network are much faster, we can use a static limit where the network structure is essentially constant. 
When the dynamics on the network are much slower, referred to as the annealed limit, then the dynamical variables effectively experience the average of the network structure over time. Then we can aggregate many snapshots regardless of their chronological order. 
In between these two limits, neither type of dynamics can be neglected~\cite{st-onge_phase_2018}. Then it is less clear how or if the history of the network's structure can be compressed while remaining faithful to the dynamics on the network.

Here, we quantify the importance of chronology in a sequence of network snapshots by considering its effects on dynamics. Our goal is to compress unimportant structural changes while preserving changes that significantly affect the dynamical process. By compression, we mean a reduction in the number of snapshots used to represent a network’s history, rather than a minimization of the description length or memory needed for these snapshots \cite{peixoto_modelling_2017}. This task has practical benefits, such as reducing the computational cost of simulation, and we consider it a fundamental scientific question: to what extent do we need to keep track of the chronological changes to a network’s structure in order to model dynamical processes on it?

We propose a method to do so by assessing the sensitivity of the dynamics to aggregating pairs of network snapshots.
We take an epidemic spreading model as our main example, but we abstract the dynamics in a way that could represent other dynamical processes on networks (e.g., synchronization of coupled oscillators or cascading failures in power grids). Specifically, we formulate a pairwise error measure using matrix commutators that captures how aggregating snapshots affects the epidemic process, and we aggregate network snapshots as long as this error remains low.  
Using synthetic networks and real data, we find that this approach is successful at producing a compressed snapshot sequence that still mimics the dynamic behavior of the original sequence.  

\begin{figure}[b!]
    \centering
    \includegraphics[width=0.49\textwidth]{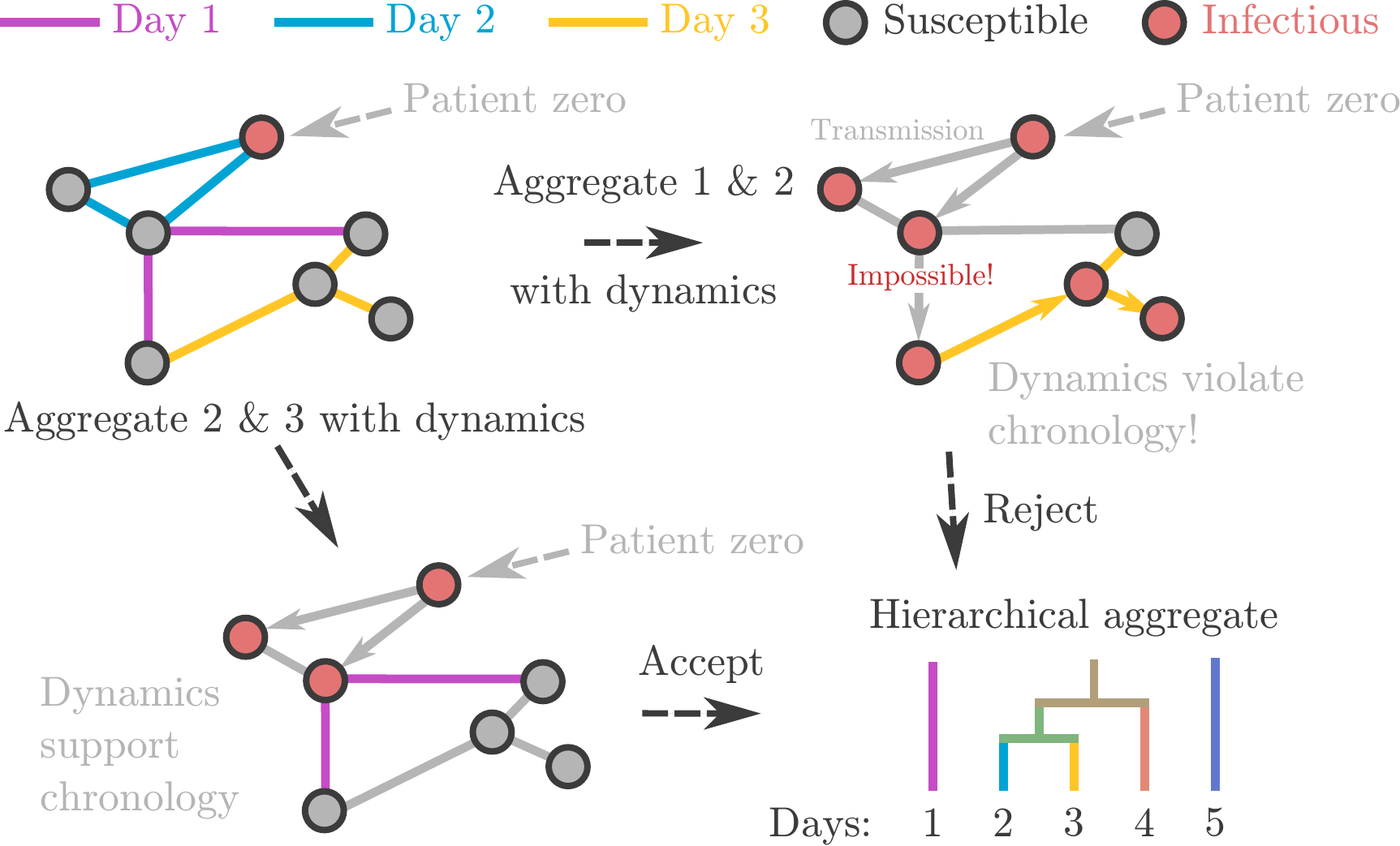}
    \caption{Schema of our hierarchical aggregation. Given network snapshots, we compare the aggregate spreading dynamics of each adjacent pair of snapshots and combine the pair with the lowest induced error, continuing until we reach a desired number of snapshots.}
    \label{fig:cartoon}
\end{figure}

\paragraph{\textbf{Analytical framework}}
We assume that we have some temporal network data consisting of a large number of snapshots of the network structure. (If we have a series of brief contacts between pairs of nodes in continuous time, this data would consist of a large series of mostly empty graphs.) We consider the dynamical error we would introduce by aggregating a given consecutive pair of snapshots into one. In an epidemic model, this error consists of creating paths that allow the contagion to progress backwards in time, as depicted in Fig.~\ref{fig:cartoon}. If aggregating two snapshots would create many such chronology-violating paths, we should keep these snapshots separate and respect their chronological order. 

In our epidemic example, a contagion spreads along edges connecting infectious nodes to their susceptible neighbors. 
We measure the effect of pairwise aggregation with a measure of error capturing the difference in the number of infected individuals, with and without aggregation over the duration of the snapshot pair. As we will see, we can approximate this error in terms of the commutator of two matrices, each of which linearizes the dynamics over the time interval of the snapshots.

\paragraph{\textbf{Linearizing the SI model}}
Let $A(t)$ be the time-varying adjacency matrix of a network with $N$ nodes. Let $P(t)$ be a vector of length $N$ where $P_i(t)$ is the probability that a node $i$ is infected at time $t$. A susceptible-infected-susceptible (SIS) model with infection rate $\beta$ and recovery rate $\gamma$ can be modeled by the following system of differential equations:
\begin{equation}\label{eqn:si_odes}
    \frac{dP_i}{dt} 
    = \beta \big( 1-P_i(t) \big) \big( A(t) \cdot P(t) \big)_i - \gamma P_i \, ,
\end{equation}
and the total number of infected individuals at time $t$ is $I(t) = |P(t)|$ where $|P|=\sum_{i=1}^N P_i$. 
In this paper we will focus on the SI model where $\gamma=0$. We then consider the early time linearization of Eq.~\eqref{eqn:si_odes} around $P=0$ (without losing much accuracy \cite{supplementary_material}), and thus have
\begin{equation}
\label{eqn:si-linearized}
    \frac{dP}{dt} = 
    \beta A(t) P(t) \, . 
\end{equation}
The solution to this equation involves a time-ordered exponential of $A(t)$. In the case of a series of snapshots $A_1,\ldots,A_\ell$, i.e., a sequence of static adjacency matrices which hold over time intervals of duration $\delta t_1,\ldots,\delta t_\ell$, we have
\begin{equation}
\label{eqn:product}
    P(t) = \left[ \prod_{i=\ell}^1 \exp\left( \beta \delta t_i A_i \right) \right] P(0)
\end{equation}
(This product is in reverse order since we treat $P(t)$ as a column vector and multiply on the left.) If all the $A_i$ commute, this product reduces to $\exp \left( \beta \sum_i \delta t_i A_i \right)$, a single exponential of their weighted average. However, in general their noncommutative nature must be taken into account.

\begin{figure}[t!]
    \centering
    \includegraphics[width=0.5\textwidth]{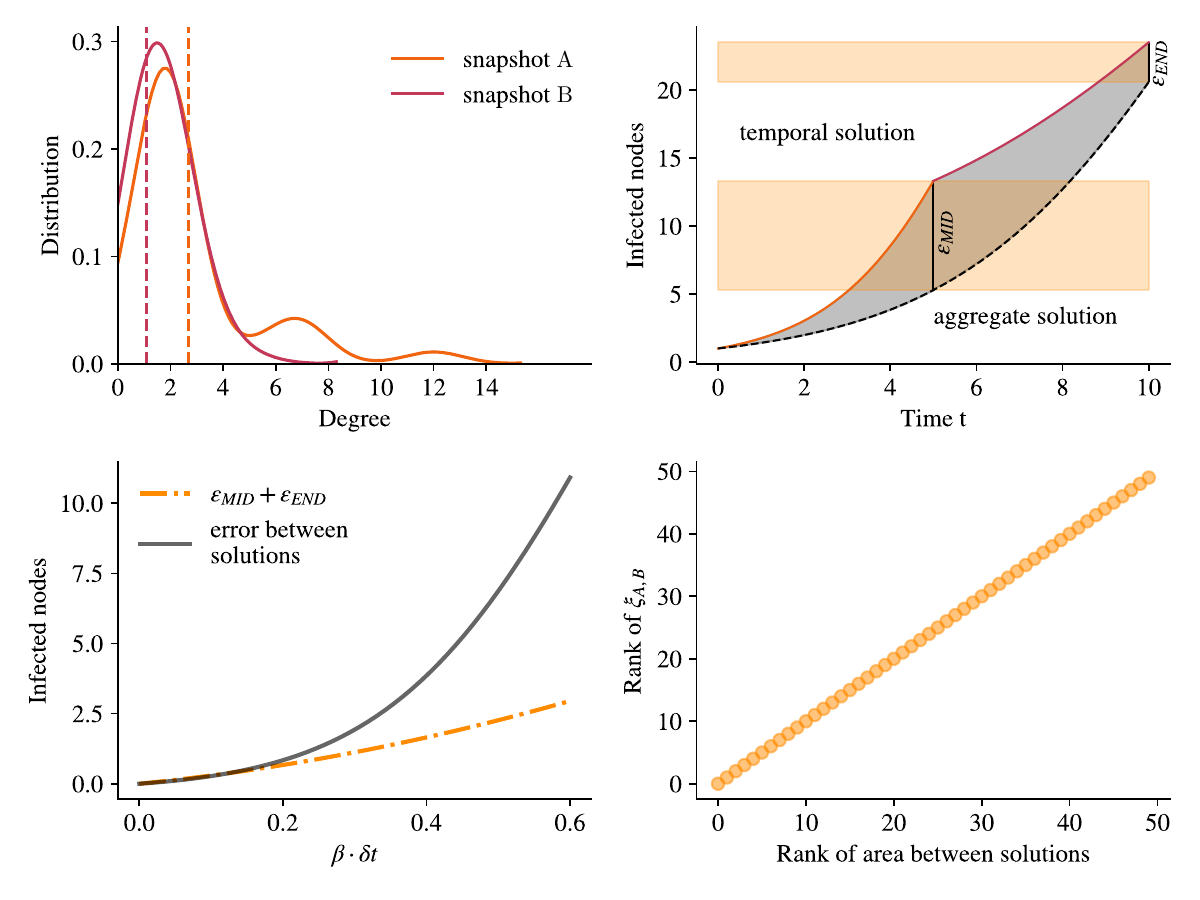}
    \caption{Top left: Degree distributions for two snapshots. Top right: ordinary differential equation solutions of the SI dynamics with $\beta =0.12$, $\delta t = 5$ on the temporal and aggregate versions of the snapshots with highlighted error terms.
    Bottom left: ODE solution difference in number of infected nodes under the temporal and aggregate regimes for varying values of $\beta t$ by varying $t=[0,5]$. Bottom right: Ranking of $\xi_{1,2}$ for snapshots 1 and 2 for increasing values of $\beta\delta_t$ compared against the integrated area between solutions.
    }
\label{fig:approximation}
\end{figure}

\paragraph{\textbf{Aggregating snapshots}}


Say we have two consecutive snapshots, $A$ and $B$, which are valid for time intervals $[t^A_0,t^A_1]$ and $[t^B_0, t^B_1]$ of duration $\delta t_A = t^A_1 - t^A_0$ and $\delta t_B = t^B_1 - t^B_0$ respectively, where $t^B_0 = t^A_1$. In the linearized SI dynamics of Eq.~\eqref{eqn:si-linearized}, 
aggregating $A$ and $B$ in a single step is equivalent to replacing the time evolution operator $\exp (\beta \delta t_B B) \exp(\beta \delta t_A A)$ with 
\[
\exp\big( \beta (\delta t_A + \delta t_B) \overline{(A,B)} \big) \, ,
\]
where $\overline{(A,B)}$ is the time-weighted average 
\begin{equation}
\label{eqn:aggregate}
    \overline{(A,B)} = \frac{\delta t_A A + \delta t_B B}{\delta t_A + \delta t_B} \, .
\end{equation}
The effect of this aggregation can be quantified through the well-known Baker-Campbell-Hausdorff formula, which expresses the  product $\exp X \exp Y$ as a single exponential $\exp Z$:
\begin{equation}
        Z 
        = X + Y + \frac{1}{2} [X,Y] + \frac{1}{12} \big(
        [X,[X,Y]]] - [Y,[X,Y]] \big) + \cdots
\end{equation}
where $[X,Y]=XY-YX$ is the matrix commutator. Setting $X=\beta \delta t_B B$ and $Y=\beta \delta t_A A$, we have
\[
\exp( \beta \delta_B B) \,\exp (\beta \delta_A A)
= \exp Z \, , 
\]
where
\begin{equation}\label{eqn:BCH}
        Z 
        = \beta (\delta t_A + \delta t_B) \overline{(A,B)} 
        + \frac{1}{2} 
        \,\beta^2 \delta t_A \delta t_B \left[ B,A \right] + O(\beta^3) \; .
\end{equation}
Up to this point, we have used a known matrix formulation of dynamical processes on networks with or without temporal aggregation. This formulation was previously used to study the impact of temporal networks on the epidemic threshold \cite{speidel2016temporal, valdano2018epidemic} and to measure auto-correlations in temporal networks \cite{valdano2018epidemic}.

Here, the correction term proportional to $[B,A]$ captures to leading order how aggregation introduces error into the dynamics. As illustrated in Fig.~\ref{fig:cartoon}, aggregation creates chronology-violating paths whenever $A$ and $B$ do not commute. For example, if $A$ includes the edge $(1,2)$ and $B$ includes the edge $(2,3)$, then a disease could spread from node 1 to 3 in the chronology $BA$ but not in $AB$. The commutator $[B,A]$ counts paths allowed by $BA$ and subtracts paths allowed by $AB$, compensating for the fact that the Taylor series of $\exp( \overline{(A,B)})$ contains terms proportional to both $AB$ and $BA$.

Thus we use the correction term in Eq.~\eqref{eqn:BCH} as a measure of the error induced by aggregation, or equivalently a measure of the  sensitivity of the dynamics to the temporal ordering of $A$ and $B$. Specifically, we use the operator norm of this term, i.e., its largest singular value, as an estimate of the error induced from the start to the end of the combined interval $[t_0^A, t_1^B]$:
\begin{equation}
\label{eqn:epsilon_end}
    \eend = \frac{1}{2} \,\beta^2 \delta t_A \delta t_B \Vert [B,A] \Vert_{\textrm{op}}
\end{equation}
The operator norm bounds the effect of aggregation on any state $P(t_0^A)$ at the beginning of the interval. 
Note that $\eend$ is small if $\beta$ is small, i.e., if the epidemic dynamics is slow compared to the dynamics of the network structure. It is also small if either $\delta t_A$ or $\delta t_B$ is small, since in that case the effect of $A$ or $B$ is close to the identity.

We are interested in the effect of aggregation on the entire history of the system, not just the final state. We show an explanation of the problem in Fig.~\ref{fig:approximation}. Thus we also consider the error induced at the boundary $t_1^A=t_0^B$ between the two intervals. To leading order we have
\[
\exp(\beta \delta t_A A) - \exp\!\left(\beta \delta t_A \overline{(A,B)} \right) 
= \beta \delta t_A \left( A - \overline{(A,B)} \right) + O(\beta^2) \, . 
\]
This term is in fact symmetric in $A$ and $B$, since
\[
\beta \delta t_A \left( A - \overline{(A,B)} \right)
= \beta 
\frac{\delta t_A \delta t_B}{\delta t_A + \delta t_B} (A-B) \, . 
\]
As before we bound the possible error induced at the midpoint as the operator norm of this term,
\begin{equation}
\label{eqn:epsilon_mid}
    \emid = \beta 
\frac{\delta t_A \delta t_B}
{\delta t_A + \delta t_B} 
\Vert A-B \Vert_{\textrm{op}} \, .
\end{equation}

We scale these terms by the cumulative duration of both snapshots---capturing the total effect that  $\emid$ and $\eend$ have on the epidemic process (as visualized in the top right panel of Fig.~\ref{fig:approximation})---which culminates in our error measure defined as
\begin{equation}\label{eqn:xi_a_b}
    \xi_{A,B} = (\eend + \emid)(\delta t_A + \delta t_B).
\end{equation} 
This measure includes both the error incurred by the overall difference between $A$ and $B$, and the extent to which their chronological order matters. If our main goal is to compute the final state, we can place a smaller weight on $\emid$ or focus entirely on $\eend$. Note that while $\eend$ is second order in $\beta$ and $\emid$ is first order, either  can dominate, e.g., if $A$ and $B$ have average degree $d$ then $\Vert [A,B] \Vert_{\textrm{op}}$ can grow as $d^2$. 
This new quantification of compression error captures structure and dynamics, allowing us to introduce a dynamic-preserving algorithm for the compression of temporal networks.

\paragraph{\textbf{Compression Algorithm}}
Given a temporal network dataset as a sequence of $M$ snapshots, we can use the framework to compress the snapshots into $M-j$ snapshots via a greedy algorithm. First, the number of desired iterations $j$ is set. For steps from 1 to $j$,
\begin{enumerate}
    \item Compute the error $\xi_{A,B}$ from Eq.~\eqref{eqn:xi_a_b} for each ordered pair $A, B$ of consecutive snapshots.
    \item Identify the pair $A^*, B^* = \text{argmin}_{A,B}(\xi_{A,B})$ to compress.
    \item Replace $A$ and $B$ with with their aggregate, $(\overline{(A,B)}$ over the union $[t_0^A,t_1^B]$ of their time intervals.
\end{enumerate}
Alternately, we can compress until $\min_{A,B} \xi_{A,B}$ reaches some threshold.

This algorithm produces a hierarchical aggregation of snapshots as portrayed in Fig.~\ref{fig:cartoon}. Periods where the network structure changes slowly or where snapshots commute become single snapshots with a long duration, while the periods with rapidly changing and noncommuting structures are preserved at a higher temporal resolution.

\begin{figure}[t!]
    \centering
    \includegraphics[width=.49\textwidth]{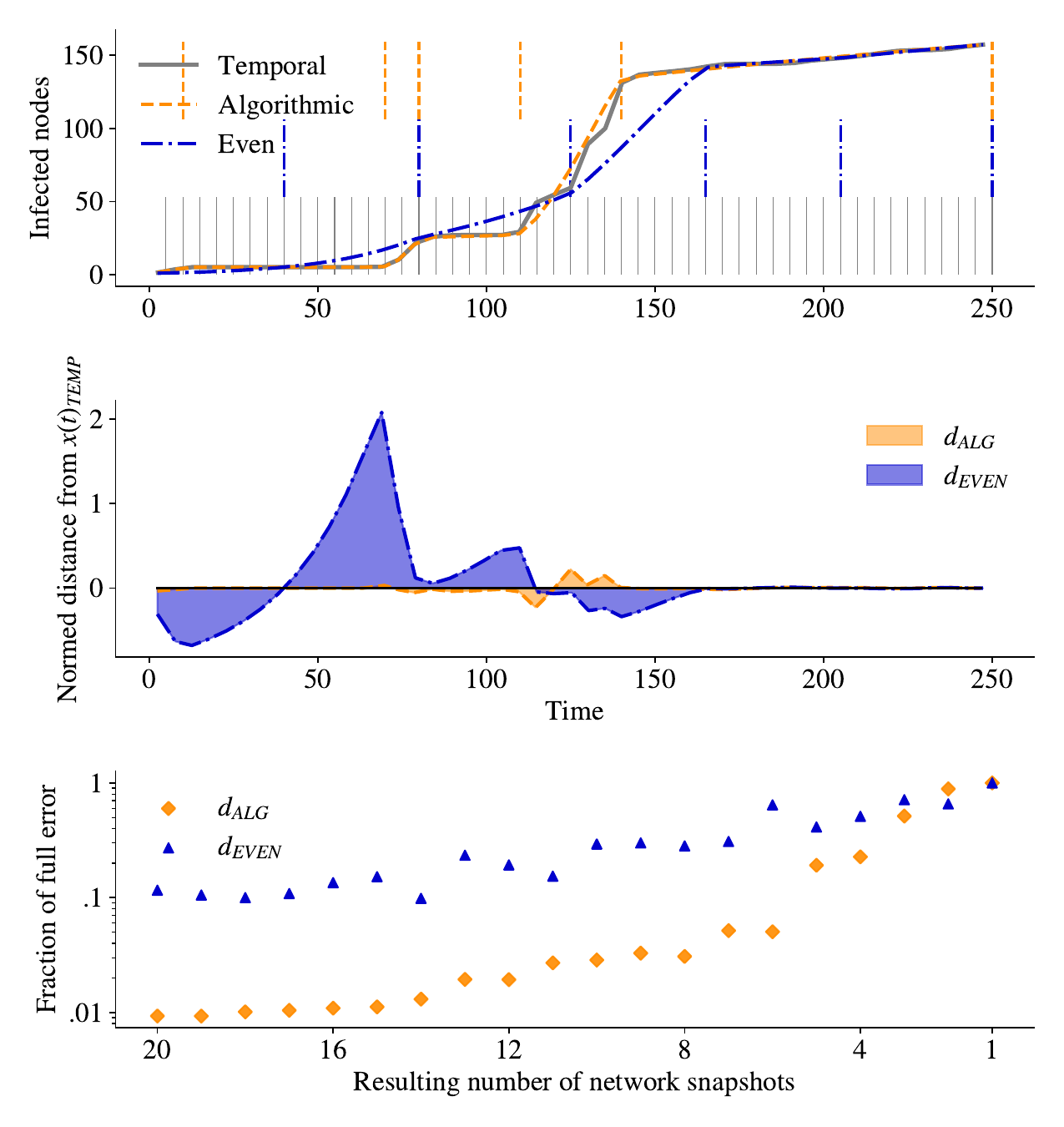}
    \caption{Compression of a series of 50 synthetic network snapshots (detailed in appendix \cite{supplementary_material}), showing the SI dynamics with $\beta=0.0017$. Top: we use our algorithm to compress the network history to $6$ snapshots of varying lengths, with boundaries shown by the orange dashed lines. We compare with the SI dynamics using even-width aggregation into $6$ windows of fixed length (blue dashed lines). The snapshots produced by our algorithm give SI dynamics closer to that on the full temporal network (gray). Middle: normalized distance from the temporal curve over time for each solution, $\dalg$ for the algorithmic solution, and $\deven$ for the even-width solution. Bottom: the vertical axis shows the shaded area from the middle panel as a function of number of aggregated snapshots, normalized by the error induced by aggregating the entire history into a single static network, measuring the error induced by aggregation as a fraction of the worst-case scenario. Generally, our algorithm results in almost ten times less error than even temporal split.
    } 
    \label{fig:synthetic}
\end{figure}

\paragraph{\textbf{Results}}

Our algorithm produces aggregated snapshots that are able to better support the epidemic dynamics on the temporal network than the standard approach of evenly dividing time into windows of some fixed length and aggregating snapshots in each window. We call $x(t)_{\textrm{TEMP}}$ the SI dynamics of Eq.~\eqref{eqn:si_odes} integrated using the full temporal dataset and $x(t)_{\textrm{ALG}}$ the dynamics integrated over the snapshots produced by our algorithm. As comparison methods, we use the common approach of evenly dividing the full time window into a certain number of snapshots of equal length (producing $x(t)_{\textrm{EVEN}}$) and an information-theoretic method that optimally compresses the full sequence as a set of modes (producing $x(t)_{\textrm{MDL}}$) \cite{kirkley2023compressing}. This last method is a parameter-free approach designed to compress general multilayer networks using the minimum description length (MDL) principle. In particular, its level of compression and error cannot be tuned as our algorithm's or the even division approach can be.

\begin{figure}[t!]
    \centering
    \includegraphics[width=.5\textwidth]{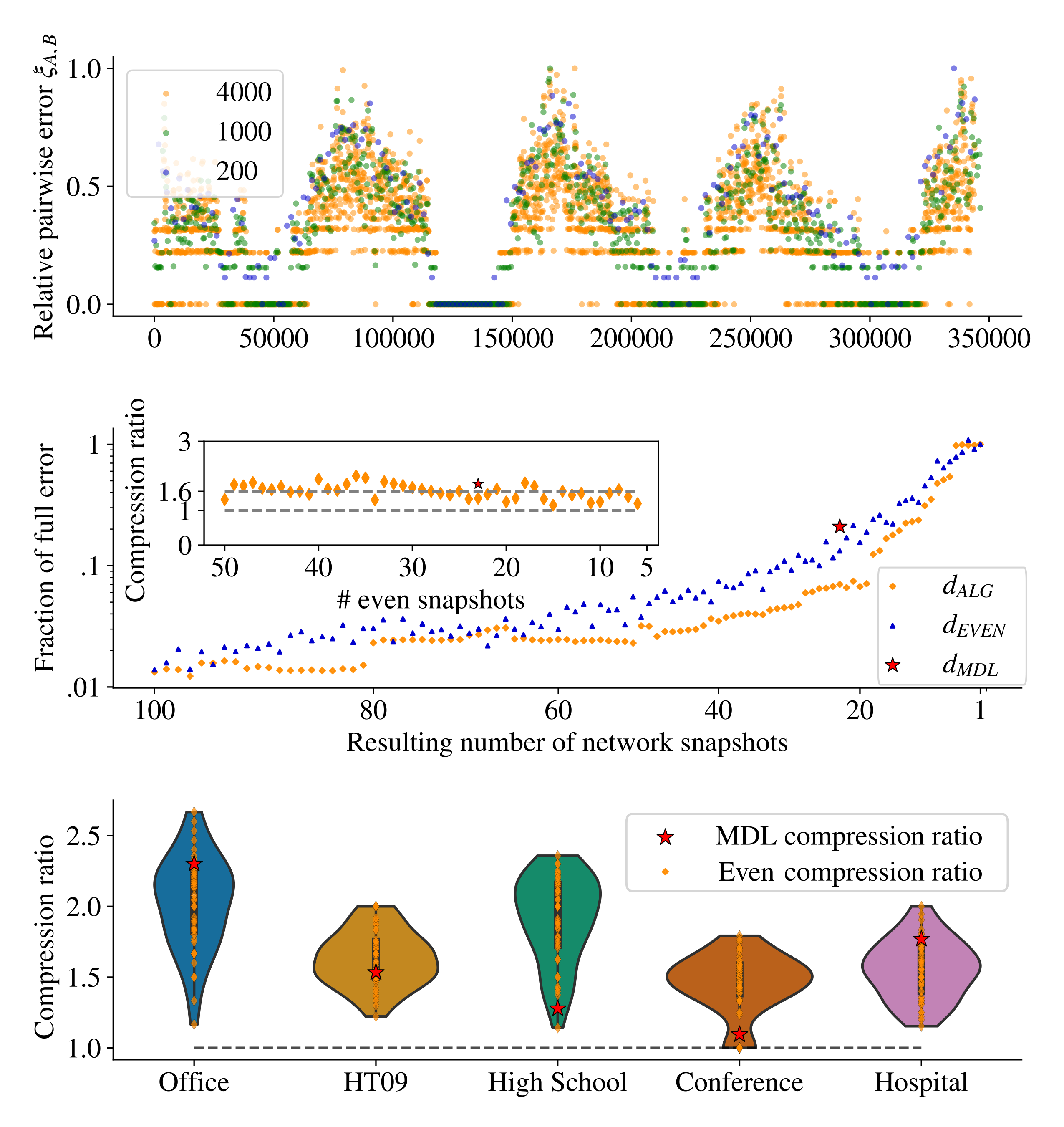}
    \caption{Application to empirical temporal network data. Top: the error measure $\xi_{S(t), S(t+1)}$ computed for consecutive snapshot pairs at 3 different levels of pre-aggregation on a hospital contact network \cite{vanhems_estimating_2013}. The hospital contact data contains contacts for approximately $9,000$ unique timestamps. We pre-aggregate by evenly coarse-graining the data to 4,000, 1,000 and 200 snapshots. Pre-aggregation of the data into small static snapshots does not affect the overall quality of our compression \cite{supplementary_material}. 
    Middle: The error of the SI process solutions, relative to the error induced by aggregating the entire history as a single static network (full error), as a function of resulting number of aggregated snapshots. The red star shows the error induced by the MDL optimal compression, which consists of 23 snapshots for this data.
    The vertical axis of the inset shows the ratio of compression achieved at a given error level by our algorithm versus even-width aggregation in orange markers, and versus MDL in a red star.
    E.g., our algorithm can compress the data to 18 snapshots while maintaining lower error than 30 even snapshots, leading to a 30/18 ($\sim 1.7$) compression ratio.
    Bottom: Summary of the inset of the middle panel across other datasets \cite{Genois2018, fournet2014, Isella:2011qo} showing the distribution of compression ratios. Our algorithm can be expected to further compress the number of snapshots by 50 to 100\%. Importantly, it always outperforms MDL compression, although the two approaches can reach very similar outputs, for example on the conference dataset.
    }
    \label{fig:hospital}
\end{figure}

We measure the multiplicative error $\dalg$ as
\begin{equation}
\label{eqn:validate_integral}
   \dalg = \int \frac{\vert x(t)_{\textrm{ALG}} - x(t)_{\textrm{TEMP}} \vert}{x(t)_{\textrm{TEMP}}} dt
\end{equation}
and similarly for $\deven$ and $\dmdl$. We apply the algorithm to synthetic networks in Fig.~\ref{fig:synthetic}, and to a dataset of four days and nights of contacts in a hospital \cite{vanhems_estimating_2013} in Fig.~\ref{fig:hospital}, along with several other temporal contact network datasets \cite{sociopatterns}.  

We show in the top panel of Fig.~\ref{fig:hospital} how the sensitivity of certain temporal ranges is maintained over a large range of resolution, which allows for pre-aggregation of data to improve the speed of the algorithm.
The error metric also allows us to identify the daily patterns of the contact data at a glance. 
Once integrated in the compression algorithm, the middle panel shows how we can aggregate over nights at the hospital, and capture the daily activity in one or two snapshots. 
As seen in Fig.~\ref{fig:synthetic} (bottom) and Fig.~\ref{fig:hospital} (middle), our algorithm compresses more effectively than even-width time windows and the MDL method. In fact, for a given number of even-width aggregation steps, the algorithm can attain the same level of error after significantly more aggregation steps ($\sim$50-200\%). We call this additional aggregation the \textit{compression ratio} and use it to summarize the results on other temporal data sets in Fig.~\ref{fig:hospital} (bottom).

\paragraph{\textbf{Discussion}}
The measure $\xi_{A,B}$ of the error induced by aggregating adjacent snapshots, and our hierarchical compression algorithm based on it, offer at least four interesting applications.

First, the approach can directly provide bounds of accuracy when studying dynamics on temporal networks with tools developed for spreading on static networks.
Second, our algorithm can help compress 
large sequences of temporal networks by aggregating  consecutive pairs of networks to reach a desired level of simplification. Our algorithm could also be modified to aggregate snapshots until the expected error caused by this aggregation is smaller than some threshold.
As shown in Fig.~\ref{fig:hospital} using real temporal interaction data, this approach allowed us to consistently meet a certain level of error while decreasing the number of required network snapshots almost by half compared to aggregating into even-width windows.

Third, the error can be used to estimate the accuracy of data collection in the first place by testing how compressible the data might be. This could help focus data collection efforts by identifying places and times with fast temporal variations, as in the top panel of Fig.~\ref{fig:hospital}. Fourth, the error can be used on non-temporal data to compare the structure of any two networks that share some of the same nodes and support the same dynamics. At its core, our approach is a network comparison tool: How different are dynamics on two networks compared to dynamics on their average?

One important limitation is the greediness of our algorithm, as it can get stuck in sub-optimal compression sequences. 
Future work should also explore how to predict the optimal stopping point of temporal compression.
We hope that our work will inspire more tools to compress temporal network data while preserving the dynamical processes they support, which is an area rich in possible applications.

\paragraph{\textbf{Acknowledgements}} A.A. and L.H.D. acknowledge support from the National Institutes of Health 1P20 GM125498-01 Centers of Biomedical Research Excellence Award and the National Science Foundation Grant No. DMS-1829826. C.M. is supported by NSF grant BIGDATA-1838251.

\end{document}